\documentclass{aa}
\usepackage{graphics}
\usepackage{amssymb}

\begin{document}

\title{Temporal aspects and frequency distributions \\ of solar soft X-ray flares}
\titlerunning{Solar soft X-ray flares}
\author{A.~Veronig\inst{1}
 \and M.~Temmer\inst{1}
 \and A.~Hanslmeier\inst{1}
 \and W.~Otruba\inst{2}
 \and M.~Messerotti\inst{3}
 }
\offprints{A. Veronig} \mail{asv@igam.uni-graz.at}

\institute{Institut f\"ur Geophysik, Astrophysik \&
 Meteorologie, Universit\"at Graz, Universit\"atsplatz 5, A-8010 Graz, Austria
 \and Sonnenobservatorium Kanzelh\"ohe, A-9521 Treffen, Austria
 \and Osservatorio Astronomico di Trieste, Via
 G. B. Tiepolo 11, I-34131 Trieste, Italy
 }

\date{Received 13 September 2001 / Accepted 15 November 2001}

\abstract{A statistical analysis of almost 50\,000 soft X-ray
(SXR) flares observed by GOES during the period 1976--2000 is presented.
On the basis of this extensive data set, statistics on
temporal properties of soft X-ray flares, such as duration, rise
and decay times with regard to the SXR flare classes is presented.
Correlations among distinct flare parameters, i.e. SXR peak flux,
fluence and characteristic times, and frequency distributions of
flare occurrence as function of the peak flux, the fluence and the
duration are derived. We discuss the results of the analysis
with respect to statistical flare models, the idea of coronal heating
by nanoflares, and elaborate on implications of the obtained results on
the Neupert effect in solar flares.
}

\maketitle

\section{Introduction}

With the availability of space-borne instrumentation, observations
of solar flare phenomena in X-rays became possible in the 1960s.
Disk-integrated soft X-ray emission measurements of the Sun have
been collected more or less continuously since 1974 by the
National Oceanic and Atmospheric Administration (NOAA), providing
an almost unbroken record fully covering solar cycles~21 and~22
and the rising phase of cycle~23 (Garcia 2000).

Statistical investigations on temporal aspects of solar flares observed in
various soft X-ray wavelengths have been carried out by Culhane \&
Phillips (1970), Drake (1971), Thomas \& Teske (1971), Phillips
(1972), Datlowe et al. (1974), and Pearce \& Harrison (1988). In the
meantime a wealth of data has accumulated, which makes worthwhile
re-investigating the temporal characteristics of soft X-ray (SXR) flares
on an extensive statistical basis. In the present analysis we make use of SXR flares
observed by GOES during 1976--2000.

Moreover, on the basis of this comprehensive data set, we calculate
frequency distributions of SXR flares as function of the peak flux,
the fluence, i.e. the integrated flux from the start to the end of a flare,
and the event duration. Frequency distributions of flare occurrence are
related to the observational expectations from different flare models.
Furthermore, they contain information about the possibility of
coronal heating by nanoflares. Frequency distributions of solar
flares from disk-integrated SXR measurements as function of the peak flux have been
carried out by Drake (1971), Lee et al. (1995) and Feldman et al. (1997).
However, only in the paper by Drake (1971) are frequency distributions
of SXR fluence measurements also presented. Shimizu (1995) used spatially
resolved observations from transient SXR brightenings and investigated frequency
distributions as function of energy.

Soft X-ray measurements are an important counterpart to
observations of flares in hard X-rays (HXR). From several
observations it is reported that the SXR light curve
has a similar shape as the time integral of the HXR curve.
This led to the idea that there is a causal relationship
between hard and soft X-ray emission of a flare, the so-called
Neupert effect (Neupert 1968; Dennis \& Zarro 1993). It supports a
flare model, known as the thick-target model (Brown 1971), in which
the HXR emission is electron-ion bremsstrahlung produced by energetic
electrons as they reach the dense layers of the chromosphere.
Only a small fraction of the electron beam energy is
lost through radiation; most of the loss is due to Coulomb collisions,
which serve to heat the ambient plasma.
Due to the rapid deposition of energy by the particle beams, the
energy cannot be radiated away at a sufficiently high rate and a
strong pressure imbalance develops, causing the heated plasma to expand
up into the corona (``chromospheric evaporation"), where this hot dense
plasma gives rise to the enhanced SXR emission (e.g., Antonucci et al. 1984;
Fisher et al. 1985; Antonucci et al. 1999, and references therein).
In this paper we make use of the flare frequency distributions and the
correlations among distinct flare parameters to infer information about
the validity of the Neupert effect in solar flares by statistical means.

The paper is structured as follows. The data set is described in Sect.~2.
In Sect.~3.1 a statistical investigation of temporal properties of SXR flares
is presented. Correlations among various flare parameters, such as characteristic
times, peak flux and fluence, are analyzed in Sect.~3.2. In Sect~3.3 frequency
distributions as function of the peak flux, the fluence and the duration are
derived. In Sect.~4 we give a summary and discussion of the main results.
Finally, the conclusions are drawn in Sect.~5.

\section{Data Set}

The present analysis is based on solar SXR flares observed by GOES
in the  0.1--0.8~nm wavelength band during the period January
1976 to December 2000. We make use of the GOES flare listing in the
Solar Geophysical Data (SGD), which contains almost 50\,000 single
events for the time span considered.

\subsection{GOES flare observations}

Since 1974 broad-band soft X-ray emission of the Sun has been
measured almost continuously by the meteorology satellites operated by
NOAA: the {\em Synchronous Meteorological Satellite} (SMS) and the {\em
Geostationary Operational Environment Satellite} (GOES).
The first GOES was launched by NASA in 1975, and the GOES series extends to
the currently operational GOES~8 and GOES~10. From 1974 to 1986
the soft X-ray records are obtained by at least one GOES-type satellite;
starting with 1983, data from two co-operating GOES are generally
available (Garcia 1994, 2000).

The X-ray sensor, part of the space environment monitor system aboard GOES,
consists of two ion chamber detectors,
which provide whole-sun X-ray fluxes in the 0.05--0.4 and 0.1--0.8~nm wavelength bands.
The initial series of satellites maintained attitude control via spin-stabilization.
GOES~8, launched in April 1994, and all subsequent GOES are three-axis
stabilized, making it possible to observe the Sun uninterrupted by spacecraft
rotations. This new operational mode facilitated an improved signal-to-background
ratio and a higher time resolution, 0.5~seconds (prior to 1994, the time
resolution was 3 seconds). Aside from the observing mode, the basic X-ray detector
is unchanged with respect to the previous instrumentation (Garcia 2000).
However, beginning with GOES~8, the dynamic range of the X-ray sensor was
extended in order to permit the most energetic SXR events to be recorded.

Solar soft X-ray flares are classified according to their peak burst intensity
measured in the 0.1--0.8 nm wavelength band by GOES. The letters (A, B, C, M, X)
denote the order of magnitude of the peak flux on a logarithmic scale, and
the number following the letter gives the multiplicative factor, i.e.,
${\rm A}n = n \times 10^{-8}$,
${\rm B}n = n \times 10^{-7}$,
${\rm C}n = n \times 10^{-6}$,
${\rm M}n = n \times 10^{-5}$, and ${\rm X}n = n \times 10^{-4}$~W~m$^{-2}$.
In general, {\it n} is given as a float number with one decimal (prior to
1980, $n$ is listed as an integer). No background subtraction is applied to
the data.

In the present statistical analysis we utilize the 1-minute average GOES data,
as listed in the SGD. The definition of the start of a GOES X-ray event
comprises the fulfillment of three conditions during four consecutive 1-minute long
intervals of observation: 1) all four values are above the B1 threshold; 2) all
four values are strictly increasing; 3) the last value is greater than 1.4 times the
value which occurred three minutes earlier. The maximum time is given by the
1-minute averaged value of the SXR peak time. The end time is
defined by the return of the flux to half the peak value above the pre-flare
level. The temporal parameters calculated in the present analysis refer to these
definitions of the start, end and maximum times of the SXR flares.

\subsection{Data description and reduction}

Table~\ref{raw classes} lists the number of flares reported for
the period January 1976 to December 2000, subdivided into the
different SXR flare classes. No class~A flares are listed,
since the SGD cover only SXR events $\ge$B1 (see also the
definition of flare onset in Sect.~2.1). As it can be
seen from Table~\ref{raw classes}, the bulk of flares belongs
to class~C ($\approx$66\%). Larger flares, i.e. M and
X, are less frequent and occur primarily during times of
maximum solar activity. Smaller flares, i.e. A and B,
actually occur more frequently than C~class flares. However, during
periods of maximum activity the X-ray background is too high to
detect A and B class flares from full-disk measurements (in extreme cases
the X-ray background may even reach M-level). Thus, the distribution
of detected flares among the SXR classes with the distinct maximum at
class~C results as an interaction of both these effects, i.e.\ the
infrequent occurrence of large flares and the restriction of
detecting small flares to periods of minimum solar activity.
The increased X-ray background during maximum solar activity may be
due to emission from many flare events as well as due to a steady coronal
heating mechanism (e.g., Feldman et al. 1997).

\begin{table}[htb]
\centering
\caption{The number of flare events for the different SXR
flare classes (B, C, M, X) and the corresponding percentage values
are listed. {\it T} denotes the total number of flares occurring in the
selected period (1976--2000).}
\begin{tabular}{crr} \hline
Class & No. events   &  No. (\%) \\  \hline
B     & 11558~~~ & 23.4~\,~ \\
C     & 32784~~~ & 66.4~\,~ \\
M     &  4708~~~ &  9.5~\,~ \\
X     &   359~~~ &  0.7~\,~ \\ \hline
{\it T}   & 49409~~~ &100.0~\,~ \\ \hline
\end{tabular}
\label{raw classes}
\end{table}

The present paper investigates temporal flare parameters as well as
characteristics of the measured SXR flux. For the temporal analysis, the
reported start, maximum and end times were checked.
Moreover, due to the fact that prior to 1997 the reported SXR flare times
were taken from the H$\alpha$ event, if there was a correlated one, also those
flares were rejected from further analysis. Applying these selection criteria,
for the temporal part of the analysis the data set was reduced to a number of 26\,745 events, covering 8\,844~B,
16\,507~C, 1\,331~M and 63~X class flares. Peak flux values are basically listed
for each event and the respective analysis makes use of the overall data set as
listed in Table~\ref{raw classes}. SXR fluence data, i.e. the integrated
flux from the start to the end of an event, are available since January
1997, amounting to 8\,400 events.

\section{Analysis and Results}

\subsection{Statistics of temporal flare parameters}

We statistically analyzed temporal aspects of SXR flares, i.e.
the duration, rise and decay times. The temporal parameters have
been derived only from those events, which fulfilled the
selection criteria given in Sect.~2.2. Figure~1 shows the
distributions of the duration, rise and decay times considering the total
of events. Each histogram reveals a pronounced negative skewness.

\begin{figure}
\centering
\vspace*{-0.15cm}
\resizebox{\hsize}{!}{\includegraphics{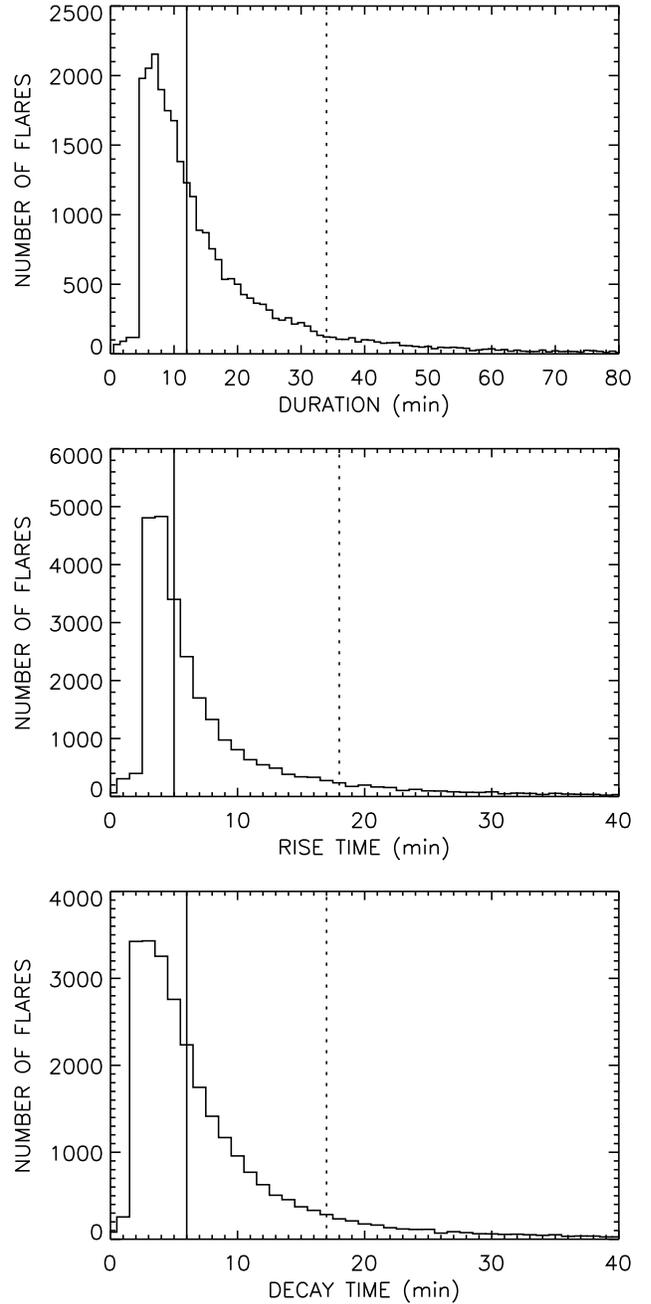}}
    \caption{Histograms of duration, rise and decay times calculated from
    the total of flares passing the temporal selection criteria. The solid line
    indicates the median value of the distribution. The dotted line
    represents the 90th percentile, which indicates the abscissa range
    covering 90\% of events.}
    \label{skew}
\end{figure}

In Table~\ref{average} we list various statistical measures
characterizing the distributions of the temporal parameters, the
arithmetic mean, the median, the mode and the 90th percentile~$P_{90}$.
The obtained results for the mode of the duration, 7.0 minutes, is
in agreement with previous studies, although different definitions
of the start and end time of SXR events are used. Phillips (1972)
and Pearce \& Harrison (1988) report a mode in the range 5--10 minutes.
Feldman et al. (1997) obtained SXR lifetime peaks in the range
6--8 minutes. Drake (1971) reports higher values, 16~minutes, considering
only flares greater than about~C2.

\begin{table}[htb]
\centering
\caption{Mean, median, mode and 90th percentile ($P_{90}$) values of the
duration, rise and decay times of the total number of flares.}
\begin{tabular}{lrrr} \hline
\multicolumn{1}{l}{Stat. measure} & \multicolumn{1}{c}{Duration} &
\multicolumn{1}{c}{Rise time} & \multicolumn{1}{c}{Decay time}\\
               & \multicolumn{1}{c}{(min)}   & \multicolumn{1}{c}{(min)}    & \multicolumn{1}{c}{(min)} \\ \hline
  Mean    & 18.2~\,~~ & 9.1~~~~  & 9.1~\,~~~~ \\
  Median  & 12.0~\,~~ & 5.0~~~~  & 6.0~\,~~~~ \\
  Mode    & 7.0~\,~~  & 4.0~~~~  & 3.0~\,~~~~ \\
$P_{90}$  & 34.0~\,~~ & 18.0~~~~ & 17.0~\,~~~~ \\  \hline
\end{tabular}
\label{average}
\end{table}

Figure~2 shows the distributions of the duration
separately for the different classes of SXR flares. It can
be seen that with increasing flare class the skewness of the
distributions decreases and the center of the distribution moves
to larger values. In Table~\ref{median} we list the median values
(plus confidence intervals) and the 90th percentile values of the
temporal parameters calculated for the different classes of SXR flares.
Since the relevant distributions are significantly asymmetric (see
Figs.~1 and~2), they are better represented by
the median, $\tilde{x}$, than by the arithmetic mean.
As a measure of statistical significance we make use of the 95\% confidence
interval, $\tilde{x} \pm c_{95}$, with
\begin{equation}\label{conf}
c_{95} = \frac{1.58~(Q_{3}-Q_{1})}{\sqrt{n}} \, .
\end{equation}
$Q_{1}$ and $Q_{3}$ denote the first and the third quartile,
respectively, $n$ the total number of data values.

\begin{figure}
\centering
\resizebox{\hsize}{18.8cm}{\includegraphics{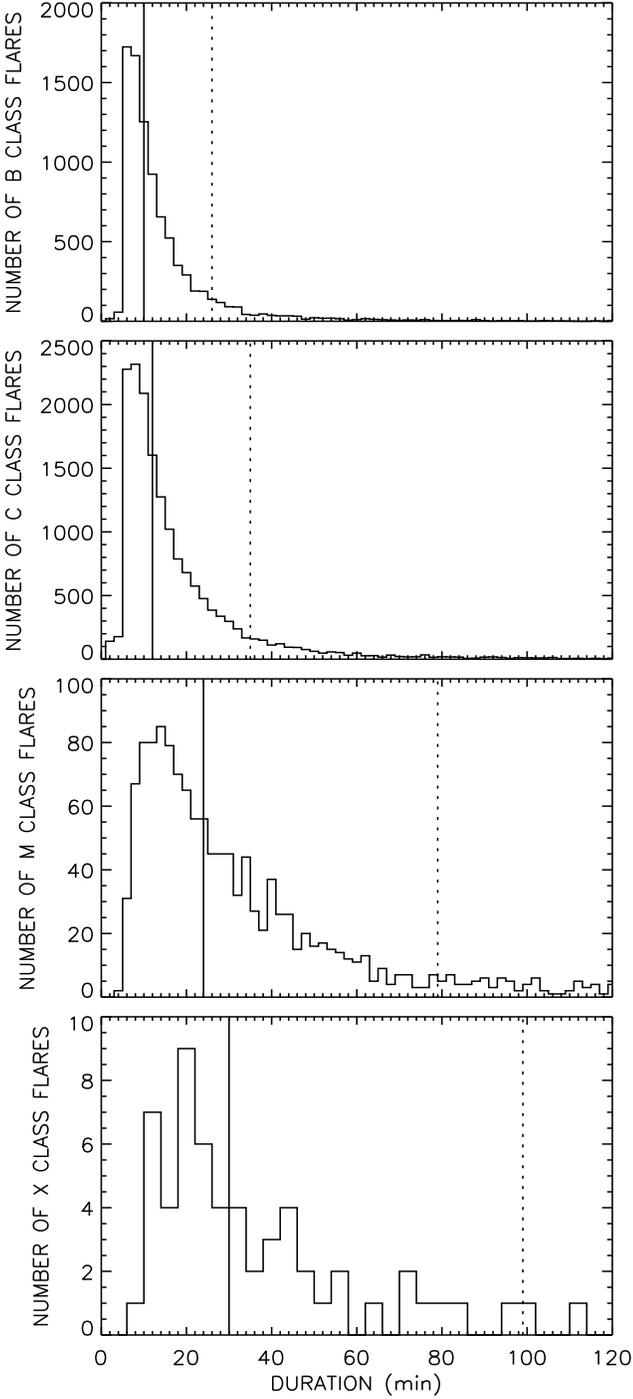}}
    \caption{Histograms of duration for the different SXR flare
    classes. The solid line indicates the median of the distribution,
    the dotted line the 90th percentile. The histograms of the
    B, C and M~class flares are represented with a bin size of
    2~minutes; for the X~class flares it is 4~minutes
    to account for the poor statistics.}
\end{figure}

\begin{table}
\centering
\caption{Median values with 95\% confidence interval, $~\tilde{x}
\pm c_{95}$, and 90th percentiles, $P_{90}$, of the duration,
rise and decay times for the different SXR classes (B, C, M, X)
and the total number of flares ({\it T}\,). All values are given in minutes.
}
\begin{tabular}
{cc@{$\pm$}c@{~}rc@{$\pm$}c@{~}rc@{$\pm$}c@{~}r} \hline
\multicolumn{1}{c}{Class} & \multicolumn{3}{c}{~~~Duration}  &
\multicolumn{3}{c}{~~Rise time} & \multicolumn{3}{c}{\,\,~Decay time}\\
        & ~~~$\tilde{x}$ & $c_{95}$& \multicolumn{1}{c}{$P_{90}$} & ~~$\tilde{x}$ & $c_{95}$& \multicolumn{1}{c}{$P_{90}$} & ~~~$\tilde{x}$ & $c_{95}$ & \multicolumn{1}{c}{$P_{90}$} \\ \hline
B       & 10.0 & 0.2 & 26.0 & ~5.0 & 0.1 & 13.0 & ~5.0 & 0.1 & 14.0 \\
C       & 12.0 & 0.2 & 35.0 & ~6.0 & 0.1 & 19.0 & ~6.0 & 0.1 & 17.0 \\
M       & 24.0 & 1.3 & 79.0 & 10.0 & 0.5 & 33.0 & 12.0 & 0.7 & 44.0 \\
X       & 30.0 & 7.4 & 99.0 & 13.0 & 2.4 & 37.0 & 14.0 & 5.2 & 89.0 \\ \hline
{\it T} & 12.0 & 0.1 & 34.0 & ~5.0 & 0.1 & 18.0 & ~6.0 & 0.1 & 17.0\\ \hline
\end{tabular}
\label{median}
\end{table}

Table~\ref{median} reveals that on average the characteristic times
increase with the flare class. The differences from one class to the
other are larger than the 95\%~confidence limits, indicating the
statistical significance of the effect. However, due to the rather
poor statistics of X class flares, the respective 95\%~confidence limits
are somewhat larger. From Table~2 and~3 it follows that the
median values of the rise and decay times are quite similar, for the
overall number of flares as well as for the different flare classes.
We want to stress that this fact is in particular related to the used
definition of the end time of an event, i.e. the return of the flux to
half the peak value above the background level at the time of the flare onset,
which obviously underestimates the decay phase. However, the actual end of an
X-ray event, i.e. the return of the coronal plasma to the state before the SXR flare,
is difficult to determine, since the background level may change during the flare
endurance and/or the decay phase may be overlaid by other events. Therefore,
definitions of the characteristic times based on the peak flux are
commonly used in statistical SXR flare studies (e.g., Culhane \& Phillips 1970;
Drake 1971; Pearce \& Harrison 1988; Lee et al. 1995).

\subsection{Correlations among flare parameters}

\begin{figure}
\centering
\resizebox{\hsize}{11.2cm}{\includegraphics{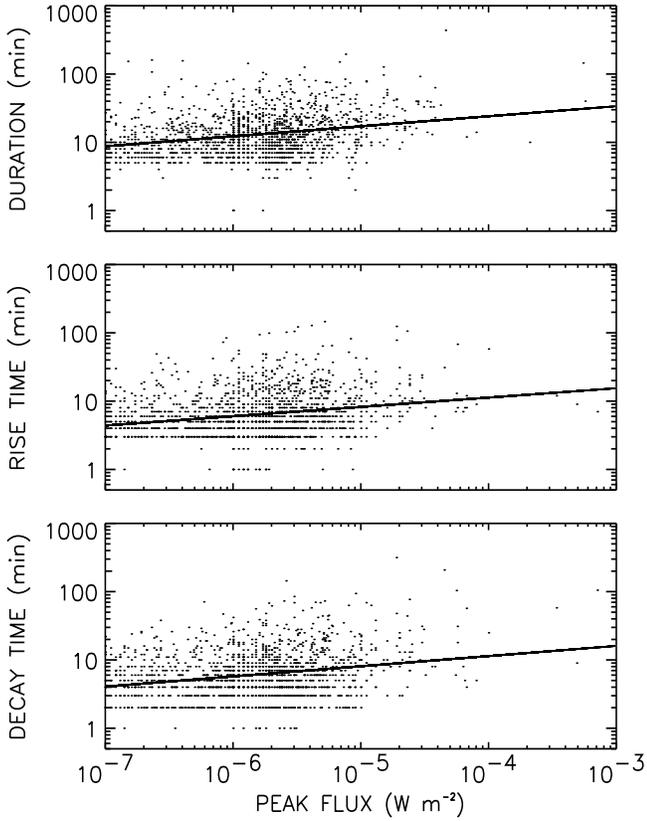}}
    \caption{Scatter plots of the flare duration (top panel), rise time
    (middle panel) and decay time (bottom panel) as function of the peak flux.
    The solid line represents the linear regression fit to the data.
    For better illustration not the whole flare sample is plotted but
    a number of 1500 randomly chosen events from the period 1976--2000.}
\end{figure}

Figure~3 shows the correlation scatter plots of the characteristic times, i.e.
duration, rise and decay times, as function of the peak flux.
The cross-correlation coefficients, calculated in log-log space,
give similar values with $r \approx 0.25$, indicating a low correlation
between the characteristic times and the peak flux.

\begin{figure}
\centering
\resizebox{\hsize}{11.2cm}{\includegraphics{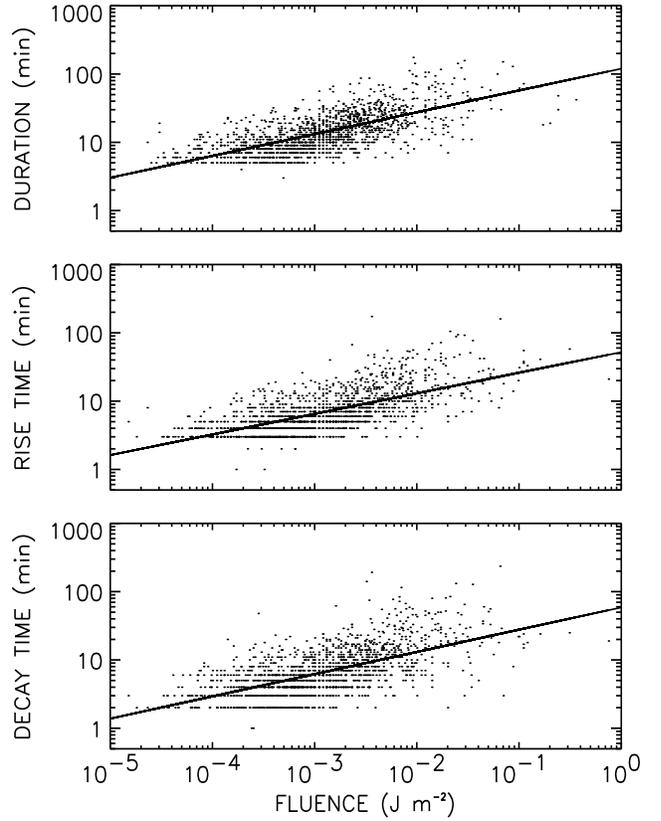}}
    \caption{Scatter plots of the flare duration (top panel), rise time
    (middle panel) and decay time (bottom panel) as function of the fluence.
     The solid line represents the linear regression fit.
     1500 randomly chosen events from the sample 1997--2000 are plotted.}
\end{figure}

In Figure~4 the scatter plots of the characteristic times as function
of the fluence are shown. The cross-correlation coefficients give $r=0.68$ for
the duration, $r=0.64$ for the rise times, and $r=0.61$ for the decay times.
The outcome that the correlation between the fluence and the characteristic
times is distinctly higher than those between the peak flux and the
characteristic times is not unexpected, since the fluence
intrinsically increases with the endurance of the event.

The top panel of Figure~5 shows the scatter plot of the fluence versus the peak
flux. The corresponding cross-correlation coefficient, $r=0.88$, is higher than
that of the fluence and the duration, $r=0.68$. From the linear fit in log-log
space we obtain the relation \mbox{${\cal F} \propto (F_{\rm P})^{1.10}$}
with ${\cal F}$ denoting the fluence and $F_{\rm P}$ the peak flux.
The bottom panel of Figure~5 shows the scatter
plot of the fluence versus the peak flux multiplied by the flare duration,
yielding $r=0.99$. From this high cross-correlation coefficient it follows
that the product of the peak flux and the duration, $F_{\rm P} \times t_{\rm dur}$,
is a good estimate of the fluence, ${\cal F}$, without accounting
for the actual time profile. From the regression analysis we obtain
the relation: ${\cal F} \propto (F_{\rm P} \times t_{\rm dur})^{0.96}$.

\begin{figure}
\centering
\resizebox{\hsize}{!}{\includegraphics{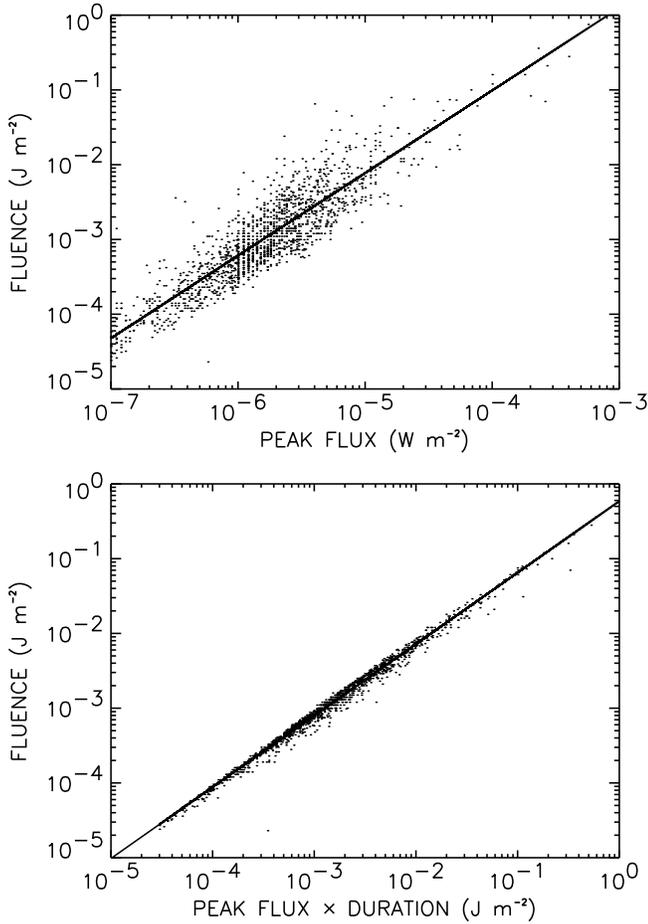}}
    \caption{Scatter plots of the fluence as function of the peak flux
    (top panel) and as function of the peak flux times duration
    (bottom panel). The solid line represents the linear regression fit.
    1500 randomly chosen events from the sample 1997--2000 are plotted.}
\end{figure}

\subsection{Frequency distributions}

Frequency distributions have been calculated for various types
of flare-associated activity, such as radio bursts, soft X-rays, hard X-rays,
interplanetary type~III bursts and interplanetary particle events
(cf. Crosby et al. 1993; Aschwanden et al. 1998; and references therein).
It has been shown that above a certain threshold (often attributed to the
sensitivity of the observations) most of these frequency distributions
can be represented by power-laws of the form
\begin{equation}
{\rm d}N = A x^{-\alpha} {\rm d}x \; ,
\label{power}
\end{equation}
where ${\rm d}N$ denotes the number of events recorded with the
parameter~$x$ of interest in the interval \mbox{$[x,x+{\rm d}x]$}. $A$ and
$\alpha$ are constants, which can be determined from a fit to the data.
For the distributions of different flare-related parameters, values in the
range \mbox{$1.4 \lesssim \alpha \lesssim 2.4$} have been obtained
(cf.\ Crosby et al. 1998). Theoretical considerations on various
aspects concerning frequency distributions of flare-related phenomena,
such as truncation effects, problems of the histogram method, time
resolution of the observations, can be found, e.g., in Lee et al. (1993),
Parnell~\& Jupp (2000) and Isliker \& Benz (2001).

\begin{figure}
\centering
\resizebox{\hsize}{!}{\includegraphics{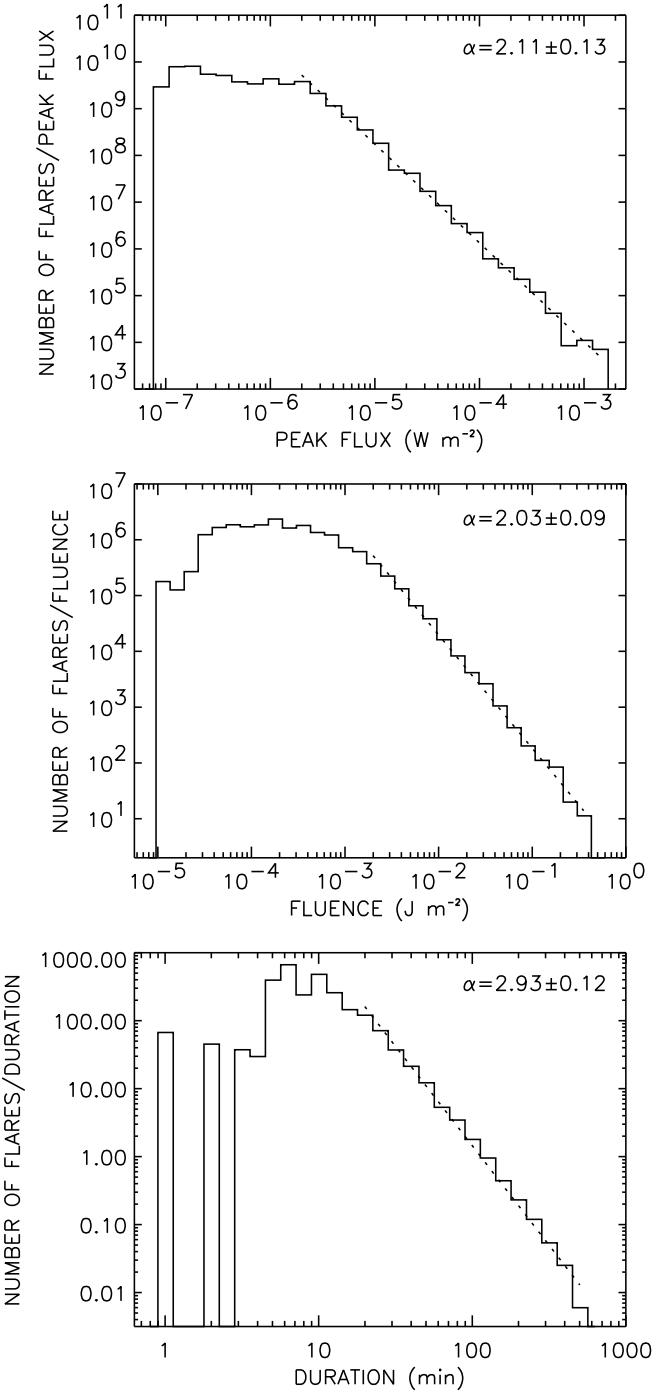}}
    \caption{Frequency distributions as function of the peak flux (top panel),
    fluence (middle panel) and duration (bottom panel). The dotted line indicates
    the least squares fit to the data.}
\end{figure}

\begin{figure}
\centering
\resizebox{\hsize}{!}{\includegraphics{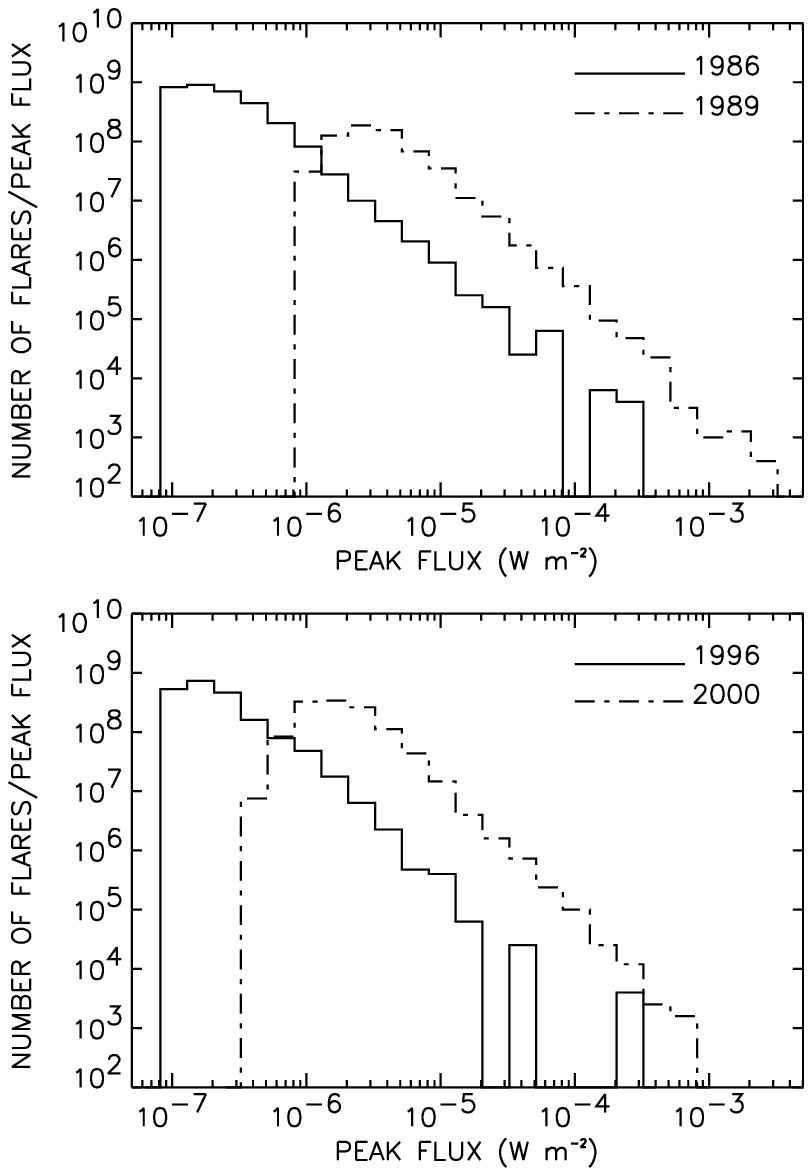}}
    \caption{Frequency distributions as function of the peak flux calculated
    separately for periods of minimum and maximum activity of solar cycle~22
    (top panel) and solar cycle~23 (bottom panel). Frequency distributions
    for times of minimum activity are represented by full lines, for times
    of maximum activity by dashed-dotted lines.}
\end{figure}

In Figure~6 we show the frequency distributions for the peak flux,
the fluence and the duration. The distributions of the
peak flux and the fluence reveal power-laws over 2--3
decades. From the slopes of the distributions in log-log space we obtain
$\alpha=2.11$$\pm$0.13 for the peak flux, and $\alpha=2.03$$\pm$0.09
for the fluence. The frequency distribution of the event
duration can also be well represented by a power-law, with
$\alpha=2.93$$\pm$0.12.

Figure~7 illustrates the frequency distributions of the peak fluxes
calculated separately for periods of minimum and maximum solar
activity. The top panel shows the respective distributions of
solar cycle~22 (calculated from the years 1986 and 1989),
the bottom panel for the current cycle~23 (1996/2000).
From the figure it can be clearly seen that during times of
minimum activity the power-law behavior extends to smaller peak fluxes,
since during these periods also less intense flares can be
detected, evidence that the turn-over of the power-law
distribution in fact is caused by the sensitivity of the
observations, and the power-law is expected to extend to even
smaller sizes. On the other hand, during times of maximum solar
activity the power-law ranges to larger peak fluxes due to the
enhanced occurrence of intense flares. However, as it can be
inferred from Figure~7, the slope of the distribution, i.e. the
power-law index, does not reveal any remarkable change in the course
of the solar cycle.

A few previous papers present frequency distributions as function of the
SXR peak flux and fluence, respectively.
Drake (1971) analyzed SXR flares measured in the 0.2--1.2~nm
wavelength range from Explorer~33 and~35, finding $\alpha=1.75$$\pm 0.10$ for the peak flux
distribution, and $\alpha=1.44\pm 0.01$ for the fluence distribution. Lee et al.
(1995), investigating the peak fluxes of SXR flares with hard X-ray counterparts
observed by GOES in the 0.1--0.8~nm band during 1981--1989, obtained $\alpha=1.86$$\pm 0.10$.
Additionally, they investigated peak count rates for the Ca~{\small XIX}
soft X-ray line emission measured by the BCS instrument aboard SMM, finding $\alpha=1.79$$\pm 0.01$.
Feldman et al. (1997) investigated GOES peak fluxes observed in the 0.1--0.8~nm
wavelength band during the period 1993--1995. By dividing the flares into domains according
to the X-ray background level, they were able to extend the power-law relationship to A1
brightness level, obtaining an average value of $\alpha=1.88$$\pm$0.21 from a
least squares fit and $\alpha=1.99$ from a non-parametric fit to the data.

It has to be noted that the value of $\alpha$ obtained in the present paper for the peak
flux distribution is somewhat larger than those of the above cited papers. This difference
might be caused by the fact that we did not apply a background subtraction, as some
other authors did (e.g., Drake 1971;  Feldman et al.\ 1997). Since the background subtraction
is relatively stronger on smaller flares than large ones, it is expected to
result in a slightly flattened frequency distribution. In the analysis of the peak
flux distributions, Drake (1971) applied a correction in order to account for the
effect that his data analysis method discriminates against small events on the basis
of the actual background level. However, no such correction was applied in the analysis
of the fluence distribution, which probably is the cause for the significantly smaller
value of the power-law index of the fluence distribution he obtained.

\section{Summary and Discussion}

\subsection{Temporal flare characteristics}

A loose correlation was found between the characteristic times and the
flare peak fluxes, $r \approx 0.25$. This outcome is also reflected in
the increasing values for the duration, rise and decay times with increasing
flare class, as listed in Table~\ref{median}. The average values
of the duration (defined as the median of the distribution) give
10~min for B, 12~min for C, 24~min for M, and 30~min for X~class
flares. Thus, on average the duration increases from B to X class
flares by a factor~3.

As it can be inferred from Table~\ref{median}, this increase is caused
in a similar manner by the rise as well as the decay times.
This differs to what was found for H$\alpha$ flares (Temmer et al. 2001),
revealing a distinctly more pronounced increase of the decay
times than of the rise times with increasing flare importance, indicating that
the cooling phase of the H$\alpha$ flare is more strongly affected
by the event strength than the phase of heating-up the plasma at the
flare site. However, we cannot exclude that a similar effect
might exist in the case of the SXR flares, since the definition of the
event end time by the return of the flux to half the peak
value possibly causes a stronger underestimation of the actual decay time
of intense events than weaker ones.

\subsection{Statistical flare models}

The distributions of the peak flux, the fluence and the duration
revealed a power-law behavior over several decades, the determined power-law indices
are \mbox{$\alpha = 2.11$$\pm$0.13}, \mbox{$\alpha = 2.03$$\pm$0.09} and
\mbox{$\alpha = 2.93$$\pm$0.12}, respectively. The first attempt to explain
the power-law distributions of flare-related phenomena
was done by Rosner \& Vaiana (1978), who developed a model based on a stochastic
flaring probability and exponential energy build-up between successive events.
However, the correlation between the strength of a flare and
the elapsed time since the previous event, predicted by the model, is
not supported by observations (Biesecker 1994; Crosby et al. 1998; Hudson et al. 1998;
Wheatland 2000).

Lu \& Hamilton (1991) were the first to propose an avalanche
model of solar flares, relating the power-law distributions
to the scale-invariant properties of a self-organized system in a critical
state. Contrary to the energy storage model, no correlation between elapsed
time and event size is expected from avalanche flare models. Moreover, since the
avalanche size distribution is insensitive to much of the microphysics, from such
a model it is also expected that the power-law distributions do not change over
the solar cycle (see, e.g., Lu \& Hamilton 1991; Lu et al. 1993).
Crosby et al. (1993) and Lu et al. (1993) reported such invariance of $\alpha$
in the course of the solar cycle for the occurrence of hard X-ray flares.
In the present analysis we calculated peak flux distributions separately
from periods of minimum and maximum solar activity (see Figure~7), which
confirms the invariance of $\alpha$ also for SXR flares. This outcome is similar to the
finding of Feldman et al. (1997) that the power-law index of the peak flux
distributions of SXR flares is insensitive to the actual SXR background level,
which is closely related to the solar cycle (e.g., Wilson 1993).

\subsection{Frequency distributions and correlations among flare parameters}

Significant correlations among the characteristic times and the fluence,
$r \approx 0.65$, were found. (Note that all cross-correlations
were calculated in log-log space.) The correlation analysis of the
fluence, ${\cal F}$, and the peak flux, $F_{\rm P}$, revealed a high
cross-correlation coefficient, $r = 0.88$. From the regression analysis it
follows \mbox{${\cal F} \propto (F_{\rm P})^{1.10}$}. It is also shown that
it is justified to estimate the fluence of an event by the product of the
peak flux and the duration, since the respective
cross-correlation coefficient amounts to $r=0.99$,
whereas ${\cal F} \propto (F_{\rm P} \times t_{\rm dur})^{0.96}$.

The obtained power-law index of the peak flux distribution, \mbox{$\alpha_{\rm P} = 2.11$},
is similar to that of the fluence distribution, \mbox{$\alpha_{\cal F} = 2.03$}. Assuming that
on average the fluence ${\cal F}$ of an event corresponds to its peak flux $F_{\rm P}$, which
can be justified by the high correlation among these parameters, then
$ \frac{{\rm d}N({\cal F})}{{\rm d}{\cal F}} \,  {\rm d}{\cal F} =
 \frac{{\rm d}N({F_{\rm P}})}{{\rm d}{F_{\rm P}}} \,  {\rm d}{F_{\rm P}}$,
where both quantities can be represented by power-law
functions with power-law index~$\alpha_{\cal F}$ and $\alpha_{\rm P}$, respectively.
On the other hand, from the scatter plot (Fig.~5, top panel) we find that ${\cal F}$ versus
$F_{\rm P}$ is well fitted by the relation ${\cal F} \propto (F_{\rm P})^s$
with \mbox{$s=1.10$}. Thus we obtain (cf. Lu et al. 1993):
\begin{equation}
{\cal F} \propto F_{\rm P}^{(1-\alpha_{\rm P})/(1-\alpha_{\cal F})} = (F_{\rm P})^s \, .
\label{al1}
\end{equation}
With this relation, the power-law index of the fluence distribution,
$\alpha_{\cal F}$, can be derived from the power-law index of the
peak flux distribution, $\alpha_{\rm P}$, and the slope of
the linear regression curve in log-log space, $s$, as:
\begin{equation}
\alpha_{\cal F} = 1 + \frac{(\alpha_{\rm P}-1)}{s} \, .
\label{al2}
\end{equation}
With $\alpha_{\rm P}=2.11$ and $s=1.10$ we obtain
$\alpha_{\cal F}=2.01$, which is very close to the value determined
from the fluence distribution, $\alpha_{\cal F}=2.03$
(cf. Fig.~6, middle panel). This means that the relation of the
power-law indices for the distribution of the peak flux and the
fluence can be interpreted in terms of the correlation between
both quantities. Since $s$, the slope obtained from the regression analysis
of the fluence versus the peak flux, is close to unity,
the value of $\alpha_{\cal F}$ is close to $\alpha_{\rm P}$; if $s$ is larger,
then $\alpha_{\cal F}$ would decrease with respect to $\alpha_{\rm P}$.

\subsection{The Neupert effect}

Several observations revealed that the shape of the SXR light curve
closely matches the time integral of the HXR emission. It has been argued that
this observation is evidence for a causal relationship between the
nonthermal hard X-ray emission and the thermal soft X-ray emission,
the so-called Neupert effect (Neupert 1968; Dennis \& Zarro 1993).
The underlying idea is that the hard \mbox{X-rays} come from
accelerated electrons impinging on coronal or chromospheric
plasma, whereas the bulk of the energy deposited by the nonthermal electrons is
converted into heating of the ambient thick-target plasma (Brown 1971).
The soft X-ray emission is due to thermal bremsstrahlung from hot dense
plasma that evaporated into the corona, as a consequence of the rapid energy
deposition. In this case, the hard X-ray emission is proportional to the time profile
of the accelerated electrons. The soft X-ray emission, which is emitted by the
plasma heated by the same nonthermal electron population, is proportional to the
accumulated energy deposited by the electrons up to a given time. Thus, it is expected
to see the Neupert effect (see McTiernan et al. 1999).

The Neupert effect, as it is commonly stated in the literature,
can be expressed as
\begin{equation}
F_{\rm P,SXR} = k \cdot {\cal F}_{\rm HXR} \, ,
\label{EqNeup}
\end{equation}
where $k$ is a proportionality factor that depends on several factors (e.g., the
magnetic field geometry, the viewing angle, etc.), and therefore may vary from flare to flare.
However, if $k$ does not depend systematically on the flare size, then the HXR fluence
and the SXR peak flux distributions should have the same shape. In particular, the
respective power-law indices should be equal, i.e.,
$\alpha_{\rm P,SXR} = \alpha_{{\cal F}{\rm ,HXR}}$ (cf. Lee et al. 1993, 1995).
However, the power-law index derived for the HXR fluence distribution,
$1.4 \lesssim \alpha_{{\cal F}{\rm ,HXR}} \lesssim 1.6$ (cf. Lee et al. 1993,
and references therein) is smaller than those of the SXR peak flux
distribution, $1.8 \lesssim \alpha_{\rm P,SXR} \lesssim 2.0$ (Drake 1971;
Lee et al. 1995; Feldman et al. 1997; this paper). It has already been pointed out
by Lee et al. (1993, 1995) that such statistical considerations lead to a discrepancy
with the Neupert effect in its simple form as stated in Eq.~\ref{EqNeup}.

A possible explanation for this discrepancy might be that the HXR and SXR emissions
are not necessarily indicative for the energies involved. As discussed in Lee et al. (1995),
the Neupert effect should exist not necessarily between the X-ray emissions but between
the energies. Namely the energy deposited by the nonthermal electrons, $\epsilon_{e^-}$,
should be equal to the maximum thermal energy contained in the plasma heated by
the same electrons, $\epsilon_{\rm th,max}$, i.e.
\begin{equation}
\epsilon_{e^-} = \epsilon_{\rm th,max} \, .
\label{NeupEn}
\end{equation}
In this case, the proportionality factor~$k$, which relates the X-ray
emissions, may be a function of the flare size, still being compatible
with the Neupert effect for the energies (Eq.~\ref{NeupEn}). One reason to expect that
$k$ indeed depends on the flare size is the finding of Feldman et al. (1996) that the
observed SXR temperature tends to increase with flare size. Therefore, the amount of
SXR emission per HXR electron may be different for large flares than for small ones
(McTiernan 2001, private communication). Moreover, as shown by McTiernan et al.
(1999), consistency of the observed HXR and SXR emission with the Neupert effect
depends on the temperature response of the SXR detector used. It has been
noted that the Neupert effect is more commonly associated with high than low temperature
SXR emission.

Another possibility, of course, is that the Neupert effect is not working for the bulk of
flares but just a subset of it. Most evidence for the Neupert effect indeed has been found
for large and impulsive flares (e.g., Dennis \& Zarro 1993; McTiernan 1999).
Applying the same considerations as in Sect.~4.3, whereas
\begin{equation}
{\cal F_{\rm HXR}} \propto (F_{\rm P, SXR})^s \, ,
\label{expr5}
\end{equation}
we find
\begin{equation}
s = \frac{(\alpha_{{\rm P,SXR}} - 1)}{(\alpha_{{\cal F}{\rm ,HXR}} - 1)} \, .
\label{EqS}
\end{equation}
Since from observations it is known that $\alpha_{{\cal F}{\rm ,HXR}} < \alpha_{\rm P,SXR}$,
it can be inferred from Eq.~\ref{EqS} that $s$ is larger than~1. Under the assumption
that the X-ray emissions are representative for the involved energies
(i.e. the proportionality factor $k$ does not depend systematically on
the flare size), $s$ is expected to be equal to~1
(cf. Eqs.~\ref{EqNeup} and~\ref{expr5}).
In this case, the found difference of the power-law indices, $\alpha_{{\cal F}{\rm ,HXR}}$
and $\alpha_{\rm P,SXR}$, indicate that the increase (decrease) of the SXR peak flux
with increasing (decreasing) HXR fluence is smaller than it would be in the case of the
Neupert effect. Any deviation from the Neupert effect basically means that the
hot plasma giving rise to the SXR emission is not heated exclusively
by the thermalization of the accelerated electrons responsible for the
HXR emission (Dennis \& Zarro 1993; Lee et al. 1993). Thus, the soft
X-ray emission might be a better indicator of the total energy released
in a flare than the HXR emission, and analogously the SXR flare
frequency distributions might better reflect the total energy
distribution of solar flares than the HXR frequency distributions
(see also Lee et al. 1993; Feldman et al. 1997).

On the other hand, if we assume that the Neupert effect formulated for the energies
(Eq.~\ref{NeupEn}) is valid for the bulk of flares, but the SXR and HXR emissions
are not directly indicative for the nonthermal and thermal energies
(i.e., $k$ may depend systematically on the flare size), then the difference of
the power-law indices of the HXR fluence and the SXR peak flux distributions
can be considered to contain information on $k$ as function of the flare size.
Since $s > 1$, and comparing Eqs.~\ref{EqNeup} and \ref{expr5}, $k$ is expected
to decrease with increasing HXR fluence or increasing SXR peak flux, respectively.
In that case, the functional dependency of $k$ on the flare size gives an
indication that the amount of SXR emission per HXR electron is smaller
for large flares than for small ones. However, in the present study we
cannot decide on these different possibilities.

\subsection{Coronal heating by nanoflares}

The actual value of the power-law index $\alpha$ determined from
various flare-related phenomena is in particular of interest with respect to
the idea of heating the corona by numerous small magnetic
reconnection events extending below the observational limit, so-called
nanoflares (Parker 1988). Hudson (1991) calculated that if the total power needed to heat
the corona is generated by flare-like events of different sizes, then the
total power is equal to the integral of event energies times their
frequency of occurrence. Assuming that the frequency of events as function of
the event energy follows a power-law of the form given in
Eq.~\ref{power} and that the energies of the largest events are much larger than
those of the smallest ones, Hudson (1991) has shown that if $\alpha < 2$, large
events dominate the total power in the distribution and nanoflares cannot
contribute much to it. Otherwise, if $\alpha \ge 2$, the more
numerous small-scale events dominate and may provide a significant contribution
to coronal heating.

However, the flare energy is not an observable quantity, and the investigations of
flare frequency distributions most often rely on peak flux or peak count rate measurements.
In a few papers though, the observed quantities have been transformed into
flare energies, and frequency distributions as function of energy
have been determined (see also Benz \& Krucker 2001).
Krucker \& Benz (1998) and Parnell \& Jupp (2000)
analyzed energy distributions of events from quiet Sun regions, so-called
network flares (Krucker et al. 1997), observed in Extreme Ultraviolet
emission, finding $\alpha$ in the range 2.3--2.6 and 2.1--2.6, respectively.
Such outcome suggests that the events at the low energy range dominate the
total power in the distribution. However, using data from the Soft X-ray
Telescope aboard Yohkoh, Shimizu (1995) studied transient brightenings
discovered in solar active regions, obtaining frequency distributions
as function of energy with $\alpha$ in the range 1.5--1.6. He estimated
that the energy provided by small events is at most 20\% of the total energy
required to heat the active region corona. Crosby et al. (1993) analyzed
frequency distributions of HXR flares. Assuming a thick-target model, they
calculated the total energy in electrons and obtained from the distribution
$\alpha \approx 1.5$.

In general, one has to be cautious in relating the frequency distributions
of observed quantities to those of flare energies, as the energy available from
flares in detail depends on their number, emission measure and temperature
(e.g., Feldman et al. 1997). Hudson (1991) made use of empirical conversion
formulae to obtain the frequency distributions of the total (radiated) flare
energy from observed peak flux measurements in various wavelengths. The
assumed proportionalities, which ensure a one-to-one correspondence
of the respective power-law indices, are criticized, e.g., by Feldman et
al. (1997). However, the fluence is a better representation of the total
energy available in flares than peak flux measurements. Under the
assumption that the fluence is proportional to the total radiated
flare energy (e.g. Krucker \& Benz, 1998), the found power-law index
of the fluence distribution, $\alpha_{\cal F}$, is also representative
of the energy distribution.

Since the determined power-law index relevant to coronal heating,
$\alpha_{\cal F} = 2.03 \pm 0.09$, is very close to the critical
value of 2, we investigated the influence of the background subtraction
on the fluence distribution, using the SXR flux just before the onset of the
flare. For the analysis we integrated this background level
over the event duration and subtracted it from the given fluence data
for the subsample January 1997 -- July 1999. As expected, from the distribution
of the background subtracted fluence data we obtain a somewhat smaller
power-law index, $\alpha_{\cal F} = 1.88 \pm 0.11$. However, within the given
error limits this value is still rather close to 2. Moreover, it has to be noted
that the derived power law index is still distinctly larger than
those obtained by Drake (1971), with $\alpha_{\cal F} = 1.44 \pm 0.01$, which
is the only other paper dealing with fluence distributions of SXR flares.

\section{Conclusions}

Frequency distributions of SXR flare occurrence as function of
the peak flux, the fluence and the event duration have been
calculated. All distributions can be described by power-law
functions over several decades. The distributions, derived
separately for the times of minimum and maximum solar activity do not
reveal any remarkable change in the power-law index, consistent
with the predictions of avalanche flare models
(e.g., Lu \& Hamilton 1991; Lu et al. 1993).

Relating the SXR fluence measurements to the total radiated flare
energy, the determined power-law index $\alpha_{\cal F}$ is also representative
of the flare energy distribution. The obtained values of $\alpha_{\cal F}$
are 2.03 for the raw fluence data and 1.88 for the background subtracted fluence data.
Both values are rather close to the critical value of 2, and no distinct conclusion
can be drawn whether small-scale events provide a significant contribution to coronal
heating or not. Moreover, deviations from the assumed proportionality between
SXR fluence and flare energies would also cause deviations from the one-to-one
correspondence of the respective power-law indices. If the slope of energy
versus fluence in log-log space is less than one, then the power-law index
of the energy distribution would be smaller than those of the fluence distribution,
and vice versa.

The power-law index of SXR peak flux distributions (see Drake 1971;
Lee et al. 1995; Feldman et al. 1997; this paper) is significantly
larger than those reported for HXR fluence distributions (cf. Lee et al. 1993,
and references therein), statistical evidence that the Neupert effect in its
commonly stated form relating the X-ray emissions (Eq.~\ref{EqNeup}) is not
valid for the bulk of flares. However, this outcome does not
necessarily mean that the Neupert effect does not work for the
more fundamental relationship between the energies (Eq.~\ref{NeupEn}).
Depending on the validity of the Neupert effect for the bulk of
solar flares, the differences in the power-law indices contain
information upon additional energy sources for the SXR-emitting
plasma, or upon the amount of SXR emission per HXR electron as a function of
the flare size. From the present analysis, based only on SXR data, we cannot
distinguish between these different possibilities. Moreover, a mixing of
both cases might exist, making a distinction even more difficult. Therefore,
a detailed statistical analysis of related SXR and HXR flares is in preparation
in order to obtain deeper insight into the Neupert effect.

\acknowledgements
The authors are very grateful to Bojan Vr\v{s}nak for the careful reading
of the manuscript and his helpful comments. A.~V. thanks Hana M\'e{s}z\'{a}rosov\'{a}
for the provision of a cross-check on the flare frequency distributions.
The authors also would like to thank James M. McTiernan for instructive
comments on the Neupert effect. A.~V., M.~T. and A.~H. acknowledge the Austrian
{\em Fonds zur F\"orderung der wissenschaftlichen Forschung} (FWF grant P13655-PHY)
for supporting this project. M. M. acknowledges the support of the
Italian Space Agency (ASI) and the Ministry for University and
Research (MURST).

\end{document}